\documentclass[runningheads]{llncs}

\usepackage[T1]{fontenc}
\usepackage{amssymb}
\def\doi#1{\href{https://doi.org/\detokenize{#1}}{\url{https://doi.org/\detokenize{#1}}}}
\usepackage{floatrow}
\newfloatcommand{capbtabbox}{table}[][\FBwidth]
\usepackage{multirow}
\usepackage{graphicx}
\usepackage{amsmath}
\usepackage{mathtools} 
\usepackage{extarrows} 
\graphicspath{{Figures/}} 
\usepackage{listings}
\begin{document}
%
\title{InvNorm: Domain Generalization for Object Detection in Gastrointestinal Endoscopy}
\titlerunning{InvNorm}
%
\author{Weichen Fan\inst{1} \and
Yuanbo Yang\inst{1} \and
Kunpeng Qiu\inst{1} \and
Shuo Wang\inst{2} \and
Yongxin Guo\inst{1}}
\authorrunning{Weichen Fan et al.}
%
\institute{National University of Singapore \and
National University of Singapore Suzhou Research Institute \\
\email{\{weichen.fan,e0787971,kunpeng\_qiu\}@u.nus.edu, shuo.w@nusri.cn, yongxin.guo@nus.edu.sg }\\
}
%
\maketitle              
\begin{abstract}
Domain Generalization is a challenging topic in computer vision, especially in Gastrointestinal Endoscopy image analysis. Due to several device limitations and ethical reasons, current open-source datasets are typically collected on a limited number of patients using the same brand of sensors. Different brands of devices and individual differences will significantly affect the model's generalizability. Therefore, to address the generalization problem in GI(Gastrointestinal) endoscopy, we propose a multi-domain GI dataset and a light, plug-in block called InvNorm(Invertible Normalization), which could achieve a better generalization performance in any structure. Previous DG(Domain Generalization) methods fail to achieve invertible transformation, which would lead to some misleading augmentation. Moreover, these models would be more likely to lead to medical ethics issues. Our method utilizes normalizing flow to achieve invertible and explainable style normalization to address the problem. The effectiveness of InvNorm is demonstrated on a wide range of tasks, including GI recognition, GI object detection, and natural image recognition.

\end{abstract}
\section{Introduction}
Gastrointestinal (GI) disease is one of the most common diseases in humans. Depending on the extent of the lesion, the GI disease could be categorized into benign GI diseases, precancerous lesions, early GI cancer, and advanced GI cancer. The early detection of GI cancer could achieve a 95\% of 5-year survival rate. The current GI disease diagnosis greatly relies on capsule endoscopy. However, the images captured by endoscopy during a single examination could be huge. Therefore, computer-assisted GI disease detection is meaningful and crucial.

\begin{figure}
\includegraphics[width=\textwidth]{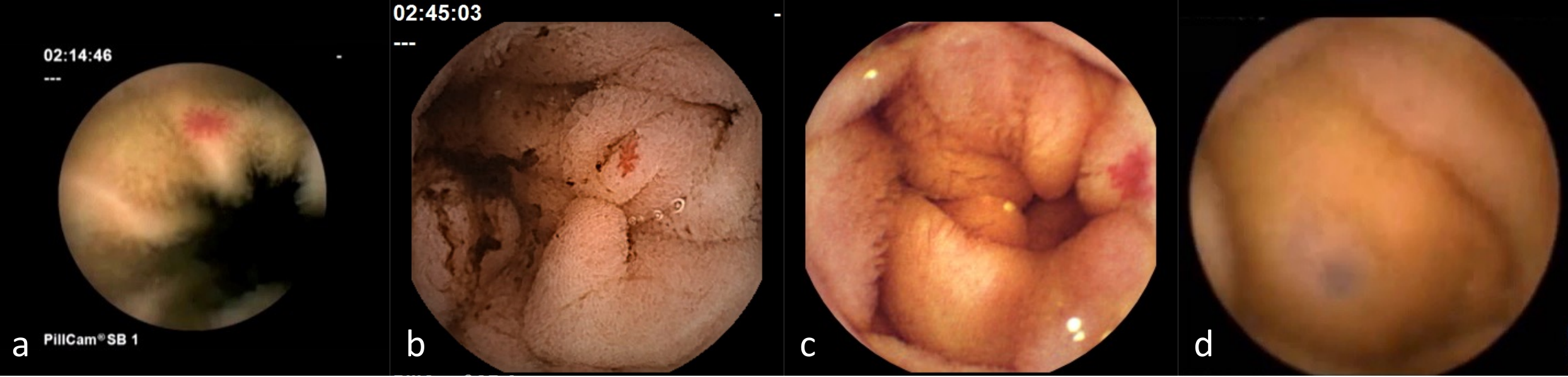}
\caption{Example of GI images in different domains.} \label{fig:intro}
\end{figure}

Due to the great success of the DL(Deep Learning) method applied in object detection, computer-assisted GI disease diagnosis has become possible. Currently, Faster-RCNN\cite{ren2015faster}, SSD\cite{liu2016ssd}, and YOLO\cite{redmon2016you} are the most common networks used in GI image analysis. Current work focuses on making traditional object detection networks perform better on GI images. However, most GI endoscopy datasets are collected with the same brand sensors from a limited number of patients. Thus, detection in GI diagnosis greatly suffers from domain gap between different endoscopies and patients, which is also a common problem in other medical images. For example, as shown in Fig. \ref{fig:intro}, vascular appeared in different patients, captured by different sensors, have a visual difference that cannot be ignored, which could be caused by illumination, camera movement, sensors' resolution, and individual patient differences.

However, current DG methods are neither too heavy(too many params) nor fail to achieve explainable and meaningful generalization, which would not be a problem in natural images, but would lead to some medical ethical issues. In this work, we propose a light, plug-in block called InvNorm(Invertible Normalization) to address the Domain Generalization problem in Gastrointestinal Endoscopy detection. Further details will be discussed in section 3.

To evaluate our method in GI detection, we further propose a multi-domain GI disease dataset consisting of 10 domains and four lesions. We apply our method to GI disease object detection in our dataset and other standard natural images DG benchmarks. Plausible results indicate the effectiveness and significance of designed modules in our method. Our contributions could be summarized as follow:\\
\indent \textbf{$\cdot$}  We propose InvNorm, a novel plug-in DG block, to solve DG problems in GI object detection.\\
\indent \textbf{$\cdot$} We propose a multi-domain GI object detection dataset.\\
\indent \textbf{$\cdot$} We demonstrate that InvNorm outperforms previous GI object detection methods and achieves SOTA performance in natural images benchmark compared with other DG methods.
\section{Related Work}
\subsection{Domain Generalization}
Most statistical learning methods assume that the source and target data are independent and identically distributed. However, the OOD(out of distribution) problem is prevalent in practice, which means the source and target domain could have a different distribution. Then DG(domain generalization) is introduced to solve the OOD problem. Given several source domains, DG is trying to learn a model which could generalize well in unseen domains.\cite{zhou2021domain} Previous methods could be categorized into data manipulation, representation learning, and learning strategy. For data manipulation, \cite{rahman2019multi,robey2021model} use generative model to improve the generalization, and \cite{shankar2018generalizing,zhou2020deep} utilize augmentation methods to address this problem. For representation learning, previous methods\cite{ilse2020diva,liu2021learning} seek to disentangle the feature into domain-invariant and domain-specific representation. For learning strategy, there are mainly three types: Meta Learning\cite{li2018learning}, Ensemble Learning\cite{mancini2018best}, and Regularization methods\cite{mixstyle,zhang2017mixup,verma2019manifold,ghiasi2018dropblock,yun2019cutmix}. Among all, Regularization methods are our main competitors, as our model is designed to be light and easy to use in any backbones for GI detection.
\subsection{Object Detection in Gastrointestinal Endoscopy}
2D Object detection in GI endoscopy is mainly inspired by Faster-RCNN\cite{ren2015faster}, SSD\cite{liu2016ssd}, and YOLO\cite{redmon2016you}. Shin et al.\cite{shin2018automatic} first proposed a model that combines Faster R-CNN with ResNet to detect colonic polys in images or videos. Tajbakhsh et al.\cite{tajbakhsh2015automatic} proposed a 3-way image representation with three CNNs for polyps detection. Bernal et al. \cite{bernal2017comparative}shows that hybrid-model could improve the overall performance. The previous detection method mainly focuses on better learning the representation in GI images. However, the domain gap, one of the most important GI image analysis problems, is ignored.

\subsection{Normalizing Flow}
Normalization flow is a type of generative model which utilizes a series of invertible mapping to transform from one distribution to another. The advantage of normalizing flow is that it could achieve accurate and efficient sampling and density estimation. Dinh et al. first propose a flow-based generative model, NICE\cite{dinh2014nice}. After that, several work\cite{kingma2018glow,dinh2016density} are proposed to improve the sample efficiency and density estimation performance. 
\section{Methodology}
\subsection{Motivation}
In DG, given n source(training) domains $S_{train}=\{S^i|i=1,..,n\}$, where $S_i$ denotes the $ith$ domains. $\{x_k,y_k\}_i$ denotes the $kth$ sample from the $ith$ domain. We assume that the joint distribution of each two domains is different. The goal of DG is to learn a robust generalizable function $f: X \xrightarrow{} Y$ from $M$ training domains, and achieve minimum error on an unseen test domain, where $P_{XY}^{test} \neq P_{XY}^i\ for\ i\in{1,..,M}$. Previous DG regularization methods try to learn domain-invariant representation or generate more representation belonging to the unseen domain to address this problem. Given a source domain $X$, the current SOTA regularization DG method MixStyle\cite{mixstyle} first map images from source domain to deep features, i.e. $E: X \xrightarrow{} F_X$. Followed the assumption proposed in AdaIN\cite{huang2017arbitrary} that the style of an image could be represented by the instance-specific mean and standard deviation of deep features. Mixstyle simply uses mixed mean and standard deviation of two deep features to replace the deep source feature, i.e., $Mixstyle: F_X \xrightarrow{} F_{mix}$. However, there exists a prerequisite, the mean and variance would represent the style only if the deep features should have a corresponding image, i.e., $D: F_{mix} \xrightarrow{} X_{mix}$, where $D$ is the decoder. If not $F_{mix}$ would be meaningless and misleading.
\begin{figure}
\includegraphics[width=\textwidth]{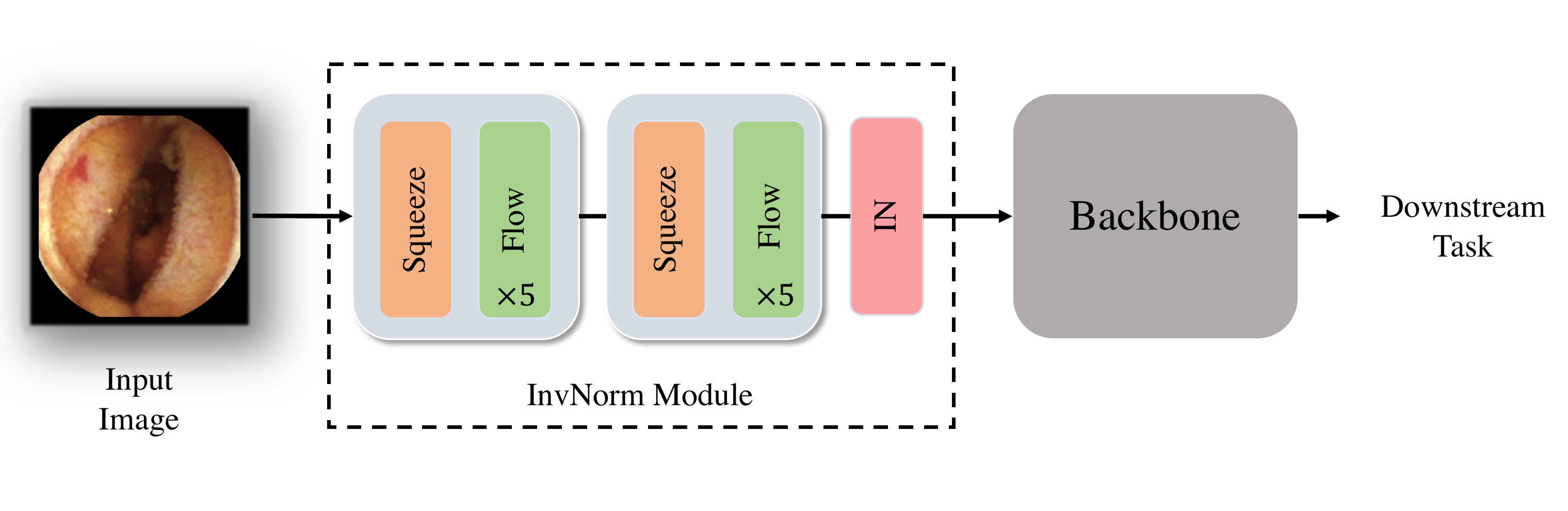}
\caption{The framework of our proposed method. The input image is first fed into our proposed InvNorm Module. The InvNorm Module consists of two blocks and an Instance Normalization Layer. Each block is structured with a squeeze layer and five flow layers. After that, any backbone is followed by the InvNorm Module for the downstream tasks.} \label{fig:method}
\end{figure}
Thus, in order to achieve explainable and meaningful style normalization, we utilize a fully-invertible normalizing flow based on GLOW\cite{kingma2018glow} as the encoder. Therefore, the encoder could map from image space to feature space, i.e. $E: X \xrightarrow{} F_X$. Then, we use an Instance normalization layer to transfer features from different source domains $F_X$ to a specific shared domain $F_S$, i.e. $IN: F_X \xrightarrow{} F_S$, and map back from feature space to image space, i.e. $E^{-1}: F_S \xrightarrow{} S$, where $S$ refers to style-normalized domain.
\subsection{Invertible Analysis}
Based on the discussion in the previous subsection, our goal is to achieve bijective mapping. \cite{kingma2018glow,dinh2014nice} propose several carefully-designed coupling and auto-regressive structures.  Behrmann\cite{behrmann2019invertible} propose an invertible ResNet under Lipschitz constrain. Followed by GLOW\cite{kingma2018glow}, our method is based on normalizing flow. To achieve bijective mapping $f: X \xleftrightarrow{} Y$ with a probability distribution $p$. Based on the change of variable theory, the bijective mapping could be written as:
\begin{align}
    P_X(x) &= P_Y(f(x))|det\frac{\partial f(x)}{\partial x}|\\
    P_Y(x) &= P_X(f(x))|det\frac{\partial f(x)}{\partial x}|^{-1}
\end{align}
where $x \in X$, $y \in Y$. Therefore, if the jocobian matrix $|det\frac{\partial f(x)}{\partial x}|$ is non-zero, the transformation could be invertible. Therefore, given a set of transformation: $f_1,f_2,...,f_n$, if $det|\frac{\partial f_i}{\partial x}| \neq 0 (i \in \{1,2,...,n\})$, then the transformation $F=f_1f_2...f_n$ would be invertible. Based on this, we would further discuss the module details in the next subsection.
\subsection{Module Detail}
\begin{figure}
\includegraphics[width=\textwidth]{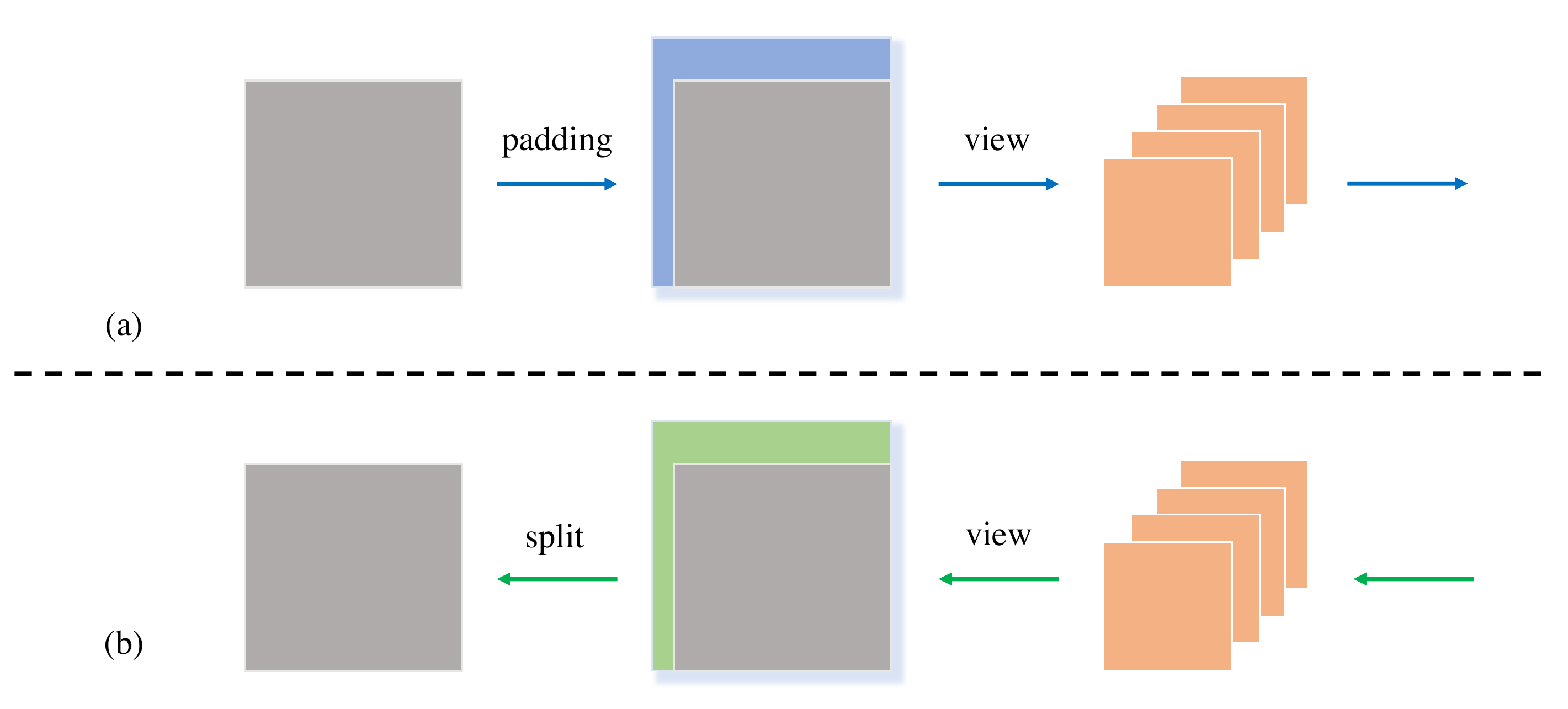}
\caption{The details of Squeeze operation. The sub-figure (a) refers to the forward pass, while sub-figure (b) refers to the backward pass. The gray box in the figure represents each channel of the tensor. The blue and green area refers to the the pixels generated by padding.} \label{fig:squeeze}
\end{figure}
Our module is based on the previous work NICE\cite{dinh2014nice}. For more details and proof, we strongly suggest readers read these papers\cite{dinh2014nice,kingma2018glow}.\\[1ex]
\noindent \textbf{A. Squeeze Operation.} As shown in Fig. \ref{fig:squeeze}, the squeeze operation consists of a padding and a view process. Unlike the previous normalizing flow methods, our model could deal with multi-scale input using the invertible squeeze operation. The input image would first be padded into a rectangle of even length and width in the forward pass. Then, each tensor would be split into four tensors along the spatial dimensions using a checkerboard pattern. Every four tensors would be merged back into one tensor along the spatial dimensions during the backward pass. After that, the padded element would be removed.\\[1ex]
\noindent \textbf{B. Flow Module.} The Flow module consists of three invertible transformations: Actnorm layer, 1x1 Convolution layer, and Coupling layer. For Actorm layer and Coupling layer, we follow the previous work GLOW\cite{kingma2018glow}. Actnorm severs as activation normalization layer. The forward and backward pass could be written as:
\begin{align}
      y_{i,j} &= s\odot x_{i,j} + b\\
      x_{i,j} &= (y_{i,j} - b) / s
\end{align}
where $i,j$ represents the spatial position of the input tensor, $s$ and $b$ refer to the learnable affine parameters.

For the Coupling layer, we follow the basic structure proposed in NICE\cite{dinh2014nice}. Given an input feature $x \in \mathbb{R}^D$, and output feature $y \in  \mathbb{R}^D$, the layer could be written as:
\begin{align}
      y_{1:d} &= x_{1:d}\\
      y_{d+1:D} &= g(x_{d+1:D}, f(x_{1:d}))
\end{align}
where $d$ splits $x$ into two partitions along channel dimension, and $g$ is the \textit{coupling function}. We define $g$ as a subtractive function in our model, and we could have:
\begin{align}
    y_{1:d} &= x_{1:d}\\
    y_{d+1:D} &= x_{d+1:D} - f(x_{1:d})
\end{align}
And the backward pass could be written as:
\begin{align}
    x_{1:d} &= y_{1:d}\\
    x_{d+1:D} &= y_{d+1:D} + f(x_{1:d})
\end{align}
we could find that $f$ has no relation to the invertibility of the coupling layer. Therefore, $f$ could be any neural network.
\\[1ex]
\noindent \textbf{C. Instance Normalization.} Different from Mixstyle\cite{mixstyle} trying to synthesize features with different samples during the training. We use an Instance Normalization\cite{ulyanov2016instance} layer followed by two normalizing flow blocks to convert all the samples to a shared style. Given an input tensor $x \in \mathbb{R}^{B\times C\times H \times W}$, where $B$ refers to batch, $C$ refers to channel, $H$ refers to height, and $W$ represents width. The IN could be formulated as:
\begin{align}
    IN(x) = \gamma \frac{x-\mu(x)}{\sigma(x)} + \beta
\end{align}
where $\gamma$ and $\beta$ are the learnable parameters, and $\mu(x)$ and $\sigma(x)$ are mean and standard deviation computed across the spatial dimension along the channel.
\section{Experiments}

\subsection{GI Object Detection}
\textbf{Dataset.}The dataset consists of four classical categories: Vascular, Ulceration, Erosion, and Bleeding, collected from ten different sensors. Most of these images are captured by PillCam from Medtronic. In order to guarantee the quality of our dataset, we got some guidelines from professional doctors. They helped clear the criteria for classifying images. Under their supervision, our medical postgraduates can accurately label the lesions in each image. In the experiment, we choose data from Sensor1, Sensor2, Sensor7, and Sensor10 as training data, and the other data will be testing data.\\[1ex]
\noindent \textbf{Implementation Details.}For object detection, we use Faster-RCNN\cite{ren2015faster} with ResNet-18\cite{he2016deep}. All the models are trained on 8 NVIDIA Tesla V100 32GB with a batch size of 128, an SGD optimizer with momentum equals 0.9, and a cosine learning rate scheduler with an initial learning rate of 0.001 for 14 epochs. For image recognition, we use ResNet-18 as a classifier to make a fair comparison with previous methods. All the classifiers are trained on a single NVIDIA Tesla V100 32GB with a batch size of 50, an SGD optimizer with momentum equals 0.9, and a cosine learning rate scheduler with an initial learning rate of 0.001 for 50 epochs. \\[1ex]
\noindent \textbf{Comparisons.}
As shown in Table. \ref{tab:sub1}, compared with the previous generalization methods, our method achieves the highest AP50 and AR(average recall) in GI object detection. As shown in Table. \ref{tab:sub2}, compared with the previous generalization methods, our method achieves the accuracy and Marco-F1 in GI image recognition.
\begin{table}
\begin{floatrow}
\centering
\capbtabbox{
\begin{tabular}{|l|l|l|}
\hline
Method &  AP50 & AR \\
\hline
Faster-RCNN & 0.158 & 0.37 \\
+ Manifold-Mixup\cite{verma2019manifold} &  0.115 & 0.29\\
+ CutMix\cite{yun2019cutmix} &  0.122 & 0.3\\
+ Mixup\cite{zhang2017mixup} & 0.142 & 0.34\\
+ DropBlock\cite{ghiasi2018dropblock} & 0.159 & 0.37\\
+ MixStyle\cite{mixstyle}& 0.171 & 0.38\\
+ \textbf{InvNorm(Ours)}& \textbf{0.217} & \textbf{0.41} \\
\hline
\end{tabular}}{\caption{The AP50 and AR compared with previous DG methods on GI object detection.}\label{tab:sub1}}

\capbtabbox{
\begin{tabular}{|l|l|l|l|l|l|}
\hline
Method &  Accuracy & Marco-F1 \\
\hline
ResNet-18 & 31.22 & 30.07 \\
+ DropBlock\cite{ghiasi2018dropblock} & 32.15 & 31.04\\
+ Manifold-Mixup\cite{verma2019manifold} &  32.26 & 28.79\\
+ CutMix\cite{yun2019cutmix} &  32.78 & 31.27\\
+ Mixup\cite{zhang2017mixup} & 33.14 & 30.2\\
+ MixStyle\cite{mixstyle}& 35.83 & 34.78\\
+ \textbf{InvNorm(Ours)}& \textbf{48.29} & \textbf{41.82} \\
\hline
\end{tabular}}{\caption{The accuracy and Marco-F1 compared with previous DG methods on GI image recognition.}\label{tab:sub2}}
\end{floatrow}
\end{table}
\subsection{Image Recognition}
\textbf{Dataset.} We choose PACS\cite{li2017deeper}, a widely used natural images classification DG benchmark for evaluation. There are seven classes, four domains, i.e., Art Painting, Cartoon, Photo, and Sketch, and in total 9991 images. Different image styles mainly cause the domain shift of this dataset.\\[1ex]
\noindent \textbf{Implementation Details.} The experiments are conducted following the leave-one-domain protocol, which means three domains are used for training and one domain for testing. We use ResNet-18\cite{he2016deep} as a classifier to make a fair comparison with previous methods. All the classifiers are trained on a single NVIDIA Tesla V100 32GB with a batch size of 50, an SGD optimizer with momentum equals 0.9, and a cosine learning rate scheduler with an initial learning rate of 0.001 for 50 epochs. \\[1ex]
\noindent \textbf{Comparisons.}
As shown in Table. \ref{tab1}, compared with the previous generalization methods, our method achieves the highest accuracy in all the categories compared with other methods by a clear margin. Besides, the InvNorm module is relatively light and easy to use. Our model combined ResNet-18 has 15.71 M params, while the original Resnet-18 has 11.69 M params.
\begin{table}
\centering
\caption{The accuracy compared with previous DG methods on PACS.}\label{tab1}
\begin{tabular}{|l|l|l|l|l|l|}
\hline
Method &  Art Painting & Cartoon & Photo & Sketch & Average\\
\hline
Manifold-Mixup\cite{verma2019manifold} &  75.6 & 70.0 & 93.4 & 65.6 & 76.2\\
CutMix\cite{yun2019cutmix} &  74.8 & 71.7 & 95.5 & 65.4 & 76.9\\
MMD-AAE\cite{li2018domain} & 75.1 & 72.4 & 95.8 & 64.3 & 76.9\\
CutOut\cite{devries2017improved} & 74.8 & 75 & 95.9 & 67.8 & 78.4\\
DropBlock\cite{ghiasi2018dropblock} & 76.3 & 75.5 & 95.8 & 69.0 & 79.2\\
Mixup\cite{zhang2017mixup} & 76.8 & 75.0 & 95.9 & 69.1 & 79.2\\
CCSA\cite{motiian2017unified} & 80.5 & 76.8 & 93.9 & 66.7 & 79.5\\
JiGEN\cite{carlucci2019domain} & 79.3 & 75.4 & 96.0 & 71.3 & 80.5\\
CrossGrad\cite{shankar2018generalizing}& 79.8 & 76.7 & 95.7 & 70.3 & 80.6\\
Epi-FCR\cite{li2019episodic}& 82.1 & 77.0 & 93.8 & 70.3 & 80.6\\
Metareg\cite{balaji2018metareg}& 83.4 & 77.1 & 95.3 & 70.1 & 81.5\\
L2A-OT\cite{zhou2020learning}& 83.3 & 78.1 & 96.0 & 73.5 & 82.7\\
MixStyle\cite{mixstyle}& 82.3 & 78.9 & 96.2 & 73.8 & 82.8\\
\textbf{InvNorm(Ours)}& \textbf{83.5} & \textbf{79.8} & \textbf{96.5} & \textbf{76.8} & \textbf{84.2}\\
\hline
\end{tabular}
\end{table}

\section{Conclusion}
This paper proposes a novel method: InvNorm, to address generalization problems in GI detection. Besides, we also provide a multi-domain dataset for GI object detection. We have proved that our method could address the non-invertible problems appeared in previous regularization DG methods. Quantitative results show that our model has achieved better performance in a DG benchmark and our GI dataset than the previous methods.
\subsubsection{Acknowledgements} This work was supported in part by NRF through the AI Singapore Programme under Award AISG-100E-2019-042.

%
%
%
\bibliographystyle{splncs04}
\bibliography{conference}
\end{document}